\documentclass[12pt]{article}

\usepackage{epsfig}         

\def\miss{\vspace{\baselineskip}}   

\def\HEAD#1
{
\miss\miss\noindent  
{\uppercase{\bf #1}}
\nopagebreak\miss
\par
}

\def\Head#1
{
\miss\noindent
{\bf #1}
\nopagebreak\miss
\par
}

\def\head#1
{
\miss
{\bf #1. \ }
}

\newenvironment{biblio}
{
\noindent
\begin{footnotesize}    
\vspace{-2.8\baselineskip}
\begin{tabbing}
\hspace{0.9cm}\=\hspace{14.7cm}\\
}
{
\end{tabbing}
\end{footnotesize}
}

\def\ref#1#2{
#1.\>\parbox[t]{14.7cm}{#2\vspace{0.1cm}}\\
}

\renewcommand{\caption}[1]
{
\begin{footnotesize}
\stepcounter{figure}
{\bf Figure \thefigure.} #1
\end{footnotesize}
}

\begin{document}

\vspace*{10\baselineskip} 
\noindent
{\bf METAL-INSULATOR TRANSITION 
\vspace{0.5\baselineskip}           
\\ IN HOMOGENEOUSLY DOPED GERMANIUM 
\vspace{0.5\baselineskip}
}\\
\par\miss\miss

\noindent
\hspace{24mm}              
\begin{minipage}{5in}
MICHIO WATANABE\\
\par
\vspace{4pt}
\noindent
\begin{tabbing}
Department of Applied Physics and Physico-Informatics, \\
Keio University, 3-14-1 Hiyoshi, Kohoku-ku, \\
Yokohama 223-8522, Japan 
\end{tabbing}
\end{minipage}
\miss



\HEAD{Introduction}
The metal-insulator transition (MIT) in doped semiconductors 
with a random distribution of impurities is 
a unique quantum phase transition in the sense that both disorder 
and electron-electron interaction play a key role (see for example 
Refs.~1 and 2).  
The metallic phase of the transition is characterized by 
a finite electrical conductivity at $T=0$, while the conductivity 
in the insulating phase vanishes in the limit of zero temperature.  
From a theoretical point of view, the correlation length in 
the metallic phase and the localization length 
in the insulating phase diverge at the critical point with the 
same exponent $\nu$, i.e., they  are proportional 
to $|N/N_c-1|^{-\nu}$ in the critical regime of the MIT, and the 
value of $\nu$ provides important information about the MIT.   
Here, $N$ is the impurity concentration and $N_c$ is 
the critical concentration for the MIT.  
Since direct experimental determination of $\nu$ is extremely 
difficult, researchers have usually determined, instead of $\nu$, 
the value of the conductivity critical exponent $\mu$ defined by 
\begin{equation}
\label{eq:critM}
\sigma (0) = \sigma^* (N/N_c - 1)^\mu 
\end{equation} 
immediately above $N_c$ ($0<N/N_c-1\ll1$).  
Here, $\sigma(0)$ is the conductivity extrapolated 
to $T=0$ and $\sigma^*$ is a prefactor.  
Values of $\nu$ are then obtained assuming the relation 
$\nu=\mu$~[3] which is valid for three-dimensional 
systems without electron-electron interaction.  
With a number of nominally uncompensated semiconductors  
$\mu\approx0.5$ has been obtained~[2].  
One of the best studied nominally uncompensated semiconductors is Si:P.  
Rosenbaum {\it et al.} studied both the doping-induced MIT 
and the uniaxial-stress-induced MIT, and showed that Eq.~(1) 
describes $\sigma(0)$ of Si:P with a single exponent $\mu\approx0.5$ 
over a wide range $0.001 \leq N/N_c-1<1$~[4].  
Here, uniaxial stress was used for ``tuning" because 
fine control of $N/N_c-1$ is difficult.   
In the same material, however, $\mu\approx1.3$ was claimed about ten years
later for a narrow regime $N_c<N<1.1N_c$ by a different group~[5].  
Moreoever, $\mu\approx1$ was reported recently 
for the MIT driven by uniaxial stress~[6].  
As the origin of the discrepancy, inhomogeneous distribution 
of the impurities has been pointed out~[7].  

For the case of melt- (or metallurgically) doped samples, 
which have been employed in most of the previous 
studies including Refs.~$4-6$,
the spatial fluctuation of $N$ due to dopant striations 
and segregation can easily be on the order of 1\% across 
a typical sample for the four-point resistance measurement 
that has a length of $\sim$5~mm or larger (see for example 
Ref.~8).  
For this reason, it will not be meaningful to discuss 
physical properties in the critical regime (e.g., 
$|N/N_c-1| < 0.01$), unless one evaluates  
the macroscopic inhomogeneity in the samples and 
its influence on the results.
In order to rule out the ambiguity arising from the inhomogeneity, 
we prepared $^{70}$Ge:Ga samples by neutron-transmutation doping (NTD) 
of isotopically enriched $^{70}$Ge single crystals.  The NTD method 
inherently guarantees the random distribution of the dopants 
down to the atomic level~[9,10].  
We show from the conductivity measurements at $T=0.02-1$~K 
that $\sigma(0)$ of the NTD $^{70}$Ge:Ga samples is described 
by Eq.~(1) with $\mu=0.50\pm0.04$ over a wide range 
$4\times10^{-4}\leq N/N_c-1<0.4$~[11].

In order to determine $\nu$ without assuming $\mu=\nu$, 
we have analyzed the temperature dependence of the 
conductivity on the insulating side of the MIT in the 
context of variable-range-hopping (VRH) conduction 
within the Coulomb gap~[12].  
Low-temperature ($T<0.5$~K) conductivity 
of the insulating $^{70}$Ge:Ga samples obeys 
$\sigma(T)=\tilde{\sigma}(T)\exp[-(T_0/T)^{1/2}]$, 
which is predicted by the VRH theory~[13], 
with an appropriate temperature 
dependence in the prefactor $\tilde{\sigma}(T)$. 
Magnetic field and temperature dependence of the conductivity 
of the samples are subsequently measured in order 
to determine directly the localization length $\xi$ and the impurity 
dielectric susceptibility $\chi_{\rm imp}$ as a function of $N$ 
in the context of the theory.  
This kind of determination of $\xi$ and $\chi_{\rm imp}$ was 
performed for compensated Ge:As by Ionov {\em et al.}~[14]  
They found $\xi\propto(1-N/N_c)^{-\nu}$ 
and $\chi_{\rm imp}\propto(N_c/N-1)^{-\zeta}$
with $\nu=0.60\pm0.04$ and $\zeta=1.38\pm0.07$, respectively, for 
samples having $N$ up to $0.96N_c$.  The significance of their result 
is the experimental verification of the relation $2\nu\approx\zeta$ that 
had been predicted by scaling theories~[15].  
However, the critical exponents of compensated samples are known to be 
different from those of nominally uncompensated samples~[11]. 
Therefore, our determination of $\xi$ and $\chi_{\rm imp}$ 
in nominally uncompensated samples is important.  

The previous effort to measure 
the impurity dielectric susceptibility as 
a function of $N$ has also contributed greatly. Hess {\em et al.} 
found $\zeta=1.15\pm0.15$ in nominally uncompensated 
Si:P~[16].  Since $\mu\approx0.5$ was determined 
for the same series of Si:P samples~[4], 
the relation $2\mu\approx\zeta$ was  again valid.  
Katsumoto has found $\zeta\approx2$ and $\mu\approx1$ for 
compensated Al$_{0.3}$Ga$_{0.7}$As:Si, i.e., 
again, $2\mu\approx\zeta$ applies~[17].  
Thus, in these cases the conclusion $2\nu\approx\zeta$ was reached 
indirectly, {\it by assuming} $\mu=\nu$.  
We, on the other hand, determine $\nu$ directly, 
i.e., we do not have to rely on the assumption $\mu=\nu$ in order to study 
the behavior of the localization length $\xi$ near the MIT.  

According to theories~[2] on the MIT 
which take into account both disorder and 
electron-electron interaction, the critical exponents 
do not depend on the details of the system, but depend 
only on the universality class to which the system belongs.  
Moreover, there is an inequality $\nu\geq2/3$, which 
is expected to apply generally to disordered systems 
irrespective of the presence of electron-electron 
interaction~[18].  
Hence, if one assumes $\mu=\nu$, 
which is derived for systems {\em without} electron-electron 
interaction, $\mu\approx0.5$ violates the inequality.  
This discrepancy has been known as the conductivity critical 
exponent puzzle.  Kirkpatrick and Belitz 
have claimed that there are logarithmic corrections to scaling 
in universality classes with time-reversal symmetry, 
i.e., when the external magnetic field is zero, and that  
$\mu\approx0.5$, found at $B=0$, should be interpreted as an ``effective" 
exponent which is different from a ``real" exponent satisfying 
$\mu\geq2/3$~[19].  
Therefore, comparison of $\mu$ with and without the time-reversal symmetry, 
i.e., with and without external magnetic fields becomes important.  
We study the MIT of $^{70}$Ge:Ga in magnetic fields up to $B=8$~T 
and show that $\mu$ changes from 0.5 at $B=0$ to 1.1 
at $B\geq4$~T~[20].  
The same exponent $\mu'=1.1$ is also found in the 
magnetic-field-tuned MIT for three different samples, i.e.,
\begin{equation}
\sigma(N,B,T\!\rightarrow\!0) \propto [B_c(N)-B]^{\mu'},  
\end{equation}
where $B_c(N)$ is the critical magnetic field for concentration $N$.  
Moreover, an excellent finite-temperature scaling~[2] 
\begin{equation}
\label{eq:finiteT}
\sigma(N,B,T)\propto T^xf(|B_c(N)-B|/T^y),  
\end{equation}
where $x/y$ is equivalent to $\mu'$, 
is obtained with the same value of $\mu'$.  
The phase diagram on the $(N,B)$~plane is successfully constructed, and 
we find a simple scaling rule which $\sigma(N,B, T\!\rightarrow\!0)$ 
would obey and derive from a simple mathematical argument that $\mu=\mu'$ 
as has been observed in our experiment.

\HEAD{Experiments}
\vspace{-\baselineskip}
\Head{Sample preparation}

All of the $^{70}$Ge:Ga samples were prepared 
by neutron-transmutation doping (NTD) 
of isotopically enriched $^{70}$Ge single crystals.  
The basic idea of NTD is as follows.  
Suppose that a nucleus in a crystal of a semiconductor captures 
a thermal neutron.  
After the capture, the nucleus is not necessarily stable.  
If it is stable, the element remains unchanged, but if it is not, 
it decays and transmutes into a new element which may act as a dopant.  
This is NTD.  Practically, a crystal is placed 
in a nuclear reactor which produces thermal neutrons.  
Since the neutron field produced by a reactor is large enough 
to guarantee a homogeneous flux over the crystal dimensions and 
the small capture cross section (typically 10$^{-24}$~cm$^2$)
of semiconductors for neutrons minimizes ``self-shadowing", 
NTD is known to produce the most homogeneous, perfectly random 
distribution of dopant down to the atomic level~[9]. 

As for Ge, there are five stable isotopes: $^{70}$Ge~(20.5\%), 
$^{72}$Ge~(27.4\%), $^{73}$Ge~(7.8\%), $^{74}$Ge~(36.5\%), and 
$^{76}$Ge~(7.8\%).  The numbers in the parentheses represent the 
natural abundance.  These five stable isotopes of Ge undergo 
the following nuclear reactions after capturing a thermal neutron.  
\begin{eqnarray}
^{70}_{32}{\rm Ge}\,+\,^1_0{\rm n} & \rightarrow & ^{71}_{32}{\rm Ge}\,
\rightarrow\,^{71}_{31}{\rm Ga}\;({\rm EC,\,11.2\,d})\,, \label{eq:70}\\
^{72}_{32}{\rm Ge}\,+\,^1_0{\rm n} & \rightarrow & ^{73}_{32}{\rm Ge}\,,
\label{eq:72}\\
^{73}_{32}{\rm Ge}\,+\,^1_0{\rm n} & \rightarrow & ^{74}_{32}{\rm Ge}\,,\\
^{74}_{32}{\rm Ge}\,+\,^1_0{\rm n} & \rightarrow & ^{75}_{32}{\rm Ge}\,
\rightarrow\,^{75}_{33}{\rm As}\;(\beta,\,{\rm 82.8\,min})\,, \\
^{76}_{32}{\rm Ge}\,+\,^1_0{\rm n} & \rightarrow & ^{77}_{32}{\rm Ge}\,
\rightarrow\,^{77}_{33}{\rm As}\;(\beta,\,{\rm 11.3\,hrs})\,
\rightarrow\,^{77}_{34}{\rm Se}\;(\beta,\,{\rm 38.8\,hrs})\,.
\end{eqnarray}
Here, $^1_0$n is a neutron, EC and $\beta$ denote electron capture and 
$\beta$~decay, respectively, and the time in the parentheses represents 
the half life.  Note that the natural Ge form both acceptors (Ga) 
and donors (As and Se) after NTD.  Empirically, the impurity compensation 
is known to affect the value of the critical exponent~[11].  
To avoid the impurity compensation, 
we use isotopically enriched $^{70}$Ge.  

The Czochralski grown, 
chemically very pure $^{70}$Ge crystal has isotopic composition 
[$^{70}$Ge]=96.2~at.~\% and [$^{72}$Ge]=3.8~at.~\%~[10].  
The as-grown crystal is free of dislocations, 
{\em p} type with a net electrically-active-impurity 
concentration less than 5$\times$10$^{11}$~cm$^{-3}$.
The thermal neutron irradiation was performed 
with the thermal to fast neutron ratio of $\approx$30:1.  
The small fraction of $^{72}$Ge becomes $^{73}$Ge which is stable, 
i.e., no further acceptors or donors are introduced.  
The post-NTD rapid thermal annealing at  650~$^{\circ}$C 
for 10~sec removed most of the irradiation-induced 
defects from the samples.  
The short annealing time is important in order to avoid the 
redistribution and/or 
clustering of the uniformly dispersed $^{71}$Ga acceptors.  
The concentration of the electrically active radiation defects 
measured with deep level transient spectrometry (DLTS) after the 
annealing is less than 0.1\% of the Ga concetration, 
i.e., the compensation ratio of the samples is less than 0.001.  
The dimension of most samples used for 
conductivity measurements was 6$\times$0.9$\times$0.7~mm$^3$.  
Four strips of boron-ion-implanted 
regions on a 6$\times$0.9~mm$^2$ face of each sample were 
coated with 200~nm Pd and 400~nm Au pads using a sputtering technique.  
Annealing at 300~$^{\circ}$C 
for one hour activated the implanted boron and removed the stress 
in the metal films.

The concentration of Ga acceptors after NTD is determined from 
the time of thermal-neutron irradiation.  The concentration 
is proportional to the irradiation time as long as the same irradiation 
site and the same power of a nuclear reactor are employed.  
\begin{figure}
\begin{center}
\epsfig{file=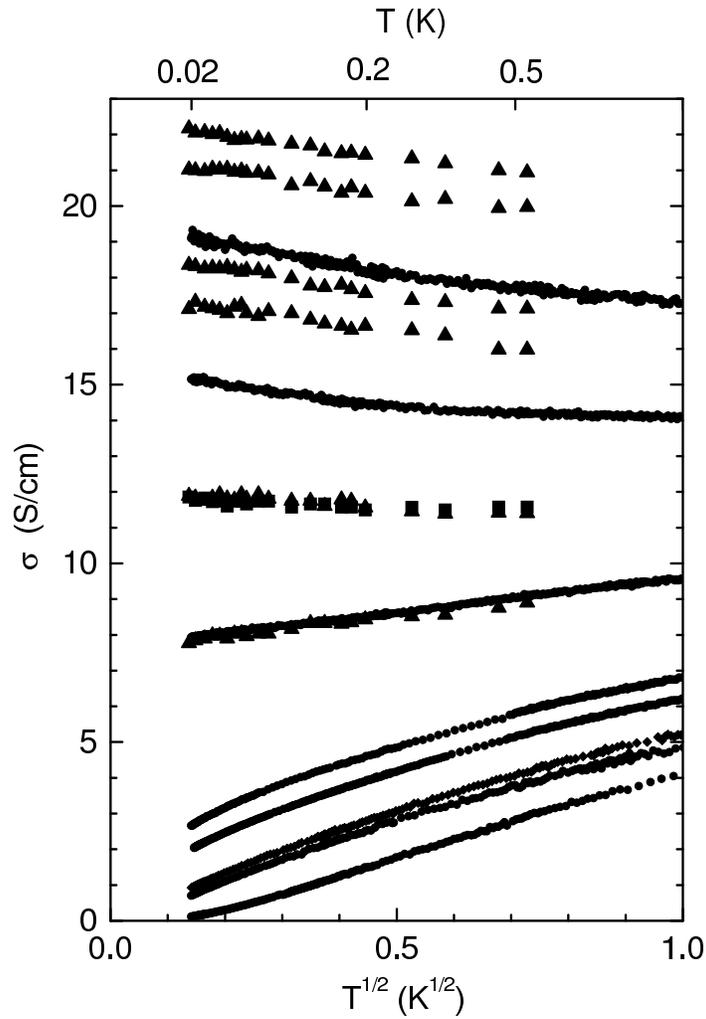,width=0.6\columnwidth,
bbllx=50,
bblly=75,
bburx=540,
bbury=785,clip=,
angle=0}
\end{center}
\caption
{
Electrical conductivity as a function of $T^{1/2}$ for NTD
$^{70}$Ge:Ga.  From bottom to top in units of
10$^{17}$~cm$^{-3}$, the Ga concentrations 
are 1.853, 1.856, 1.858, 1.861, 1.863, 1.912, 1.933, 
2.004, 2.076, 2.210, 2.219, 2.232, 2.290, 2.362, 
and 2.434, respectively.  
}
\end{figure}

\Head{Low-temperature measurements}

The electrical conductivity measurements were carried out down to 
temperatures of 20~mK using a $^{3}$He-$^{4}$He dilution refrigerator.  
All the electrical leads were low-pass filtered at the top of
the cryostat. The sample was fixed in the mixing chamber and a 
ruthenium oxide thermometer [Scientific Instrument (SI), RO600A, 
1.4$\times$1.3$\times$0.5~mm$^3$] was placed close to the sample.  
To measure the resistance of the thermometer, we used an ac resistance 
bridge (RV-Elekroniikka, AVS-47).  The thermometer was calibrated 
against 2Ce(NO$_3)_3\cdot3$Mg(NO$_3)_2\cdot$24H$_2$O (CMN) 
susceptibility and against the resistance of a canned ruthenium oxide 
thermometer (SI, RO600A2) which was calibrated 
commercially over a temperature range from 50~mK to 20~K.
We employed an ac method at 21.0~Hz to measure the resistance of 
the sample.  
The power dissipation was kept below 10$^{-14}$~W, which is 
small enough to avoid overheating of the samples.
The output voltage of the sample was detected 
by a lock-in amplifier (EG\&G Princeton Applied Research, 124A).  
All the analog instruments as well as the cryostat were placed 
inside a shielded room.
The output of the instruments was detected by digital voltmeters 
placed outside the shielded room.
All the electrical leads into the shielded room were low-pass filtered.  
The output of the voltmeters was read by a personal computer via GP-IB 
interface connected through an optical fiber.  
Magnetic fields up to 8~T were applied in the direction perpendicular 
to the current flow by means of a superconducting solenoid.

\HEAD{Results and discussion}
\vspace{-\baselineskip}
\Head{Temperature dependence of conductivity in metallic samples 
and the critical exponent for the zero-temperature conductivity}
\begin{figure}
\begin{center}
\epsfig{file=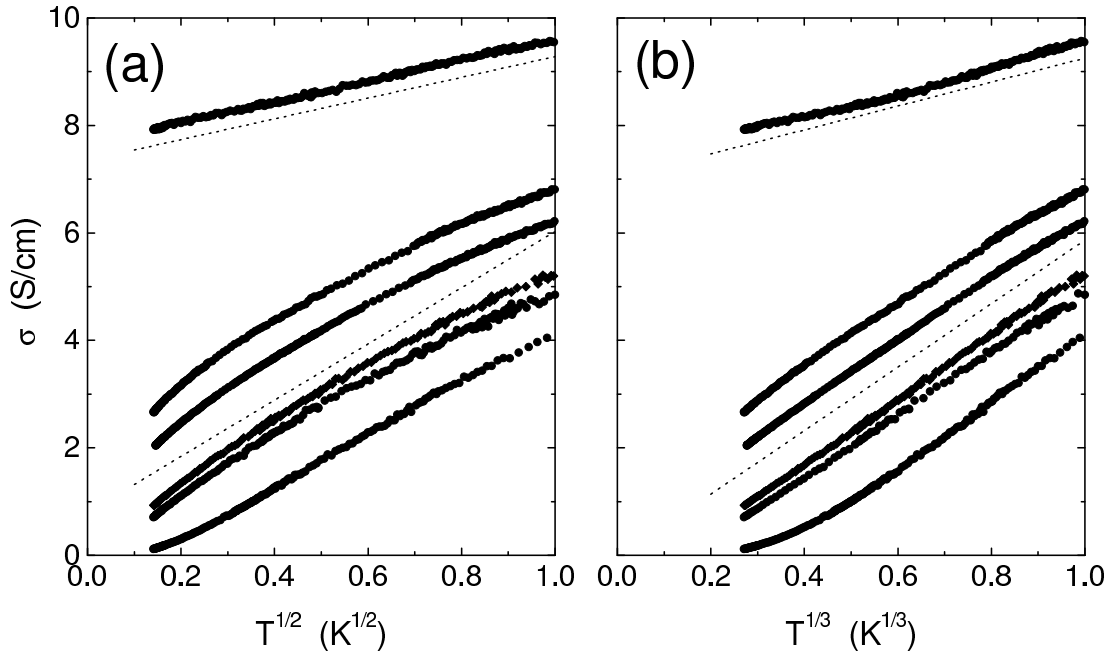,width=0.95\columnwidth,
bbllx=58,
bblly=375,
bburx=567,
bbury=680,clip=,
angle=0}
\end{center}
\caption
{
Conductivity as a function of (a)~$T^{1/2}$ and (b)~$T^{1/3}$, 
respectively, near the metal-insulator transition.
From bottom to top in units of 10$^{17}$~cm$^{-3}$, the concentrations 
are 1.853, 1.856, 1.858, 1.861, 1.863, and 1.912, respectively.
The upper and lower dotted lines in each figure 
represent the best fit using the data between 0.05 K and 0.5 K 
for the first and the third curves from the top, respectively.  
Each fit is shifted downward slightly for easier comparison.  
}
\end{figure}
The temperature dependence of the electrical conductivity mostly for 
the metallic samples 
is shown in Fig.~1.  
The temperature variation of the conductivity of disordered metal
is governed mainly by electron-electron interaction 
at low temperatures~[1], and can be written as 
\begin{equation}
\label{eq:1/2}
\Delta\sigma(T) \equiv \sigma (T) - \sigma (0) = m\sqrt{T}\,,
\end{equation}
where
\begin{equation}
\label{eq:m}
m \,=\, \frac{\;e^2}{\hbar}\frac{1}{\;4\pi^2\;}\frac{1.3}{\;\sqrt{2}\;}
\left(\frac{4}{\;3\;}-\frac{3}{\;2\;}\tilde{F} \right) 
\sqrt{\frac{k_B}{\;\hbar D\;}}\,=\, A/\sqrt{D}\,.
\end{equation}
Here, $\tilde{F}$ is a dimensionless and 
temperature-independent parameter 
characterizing the Hartree interaction
and $D$ is the diffusion constant~[1], which is related to 
the conductivity via the Einstein relation 
\begin{equation}
\label{Einstein}
\sigma =(\partial n / \partial \mu)e^2 D,
\end{equation}
where $(\partial n / \partial \mu)$ is 
the density of states at the Fermi level.  
In various reports such as Refs.~5 and 10, 
$\sigma(0)$ was obtained by extrapolating 
$\sigma(T)$ to $T=0$ assuming $\sqrt{T}$ dependence based on 
Eq.~(9).  One should note, however, that it is sound only 
in the limit of $\Delta\sigma(T) \ll \sigma(0) \approx \sigma(T)$, 
where $D$ can be considered as a constant, i.e., $m$ is constant, 
and that the inequality is no longer valid as $N$ approaches $N_c$ from 
the metallic side since $\sigma(0)$ also approaches zero.  In such cases 
$m$ in Eq.~(9) is not temperature independent and 
$\Delta\sigma(T)$ may exhibit a temperature dependence different from 
$\sqrt{T}$.  To examine this point in our experimental results, we go 
back to Fig.~1.  We see there that $\Delta\sigma(T)$ of 
the bottom five curves are not proportional to $\sqrt{T}$, while 
$\Delta\sigma(T)$ of the other higher $N$ samples are described by 
$\propto\sqrt{T}$.  The close-ups of $\sigma(T)$ for the six samples 
with positive $d\sigma/dT$ in the scale of $\sqrt{T}$ and $T^{1/3}$ are 
shown in Figs.~2(a) and 2(b), respectively.  
The upper and lower dotted lines represent the best fit 
using the data between 0.05~K and 0.5~K for the samples with 
$N=1.912\times10^{17}$~cm$^{-3}$ and 
$N=1.861\times10^{17}$~cm$^{-3}$, 
respectively.  Each fit is shifted downward slightly for the sake 
of clarity.  From this comparison, it is clear that a 
$T^{1/3}$ dependence rather than a $\sqrt{T}$ dependence 
holds for samples in the very vicinity of the MIT.   
The opposite is true for the curve at the top.  
This means that the $\sqrt{T}$ dependence in 
Eq.~(9) is replaced by a $T^{1/3}$ dependence 
as the MIT is approached.  

\begin{figure}
\begin{center}
\epsfig{file=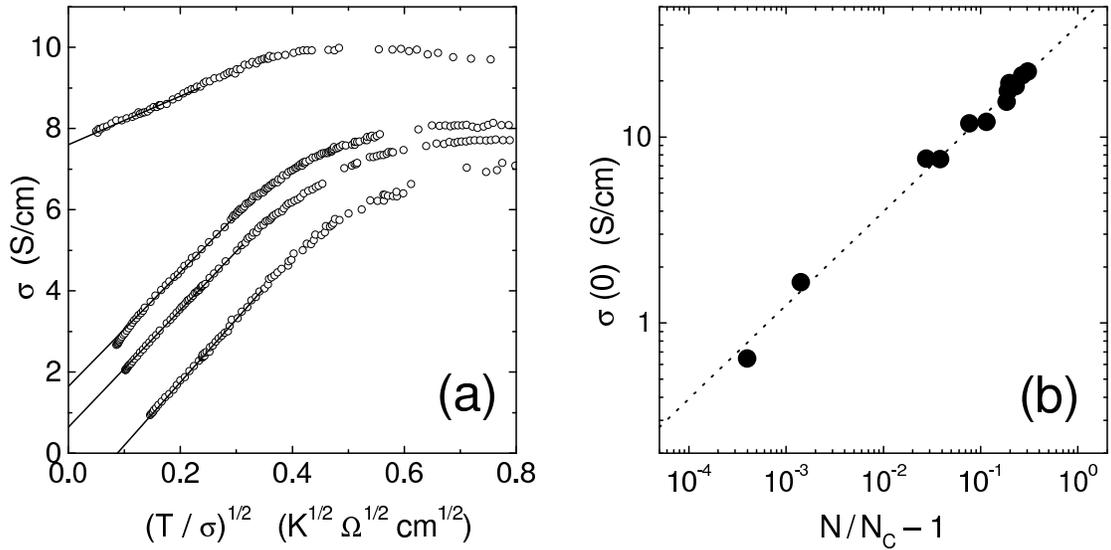,width=0.95\columnwidth,
bbllx=27,
bblly=375,
bburx=558,
bbury=638,clip=,
angle=0}
\end{center}
\caption
{
(a) Conductivity $\sigma$ as a function of $(T/\sigma)^{1/2}$. 
From bottom to top in units of 10$^{17}$~cm$^{-3}$, 
the concentrations are 1.858, 1.861, 1.863, and 1.912, 
respectively.  The solid lines denote the extrapolation 
for finding $\sigma(0)$.  
(b) Zero-temperature conductivity $\sigma (0)$ vs 
the dimensionless distance $N/N_c-1$ from the critical point
on a double logarithmic scale.
The dotted line represents the best power-law fit by  
$\sigma(0) \propto(N/N_c-1)^{\mu}$ where $\mu=0.50\pm0.04$. 
}
\end{figure}
A $T^{1/3}$ dependence close to the critical point for the MIT 
was predicted originally by Al'tshuler and Aronov~[21].  
They considered an interacting electron system with paramagnetic 
impurities, for which they obtained a single parameter 
scaling equation.  At finite temperatures, they assumed 
a scaling form for conductivity according to the scaling hypothesis: 
\begin{equation}
\label{eq:AA}
\sigma = \frac{\,e^2}{\hbar \, \xi'} \: f_1(\xi' /L_T), 
\end{equation}
where $\xi'$ is the correlation length 
and $L_T\equiv\sqrt{\hbar D/k_BT}$ is 
the thermal diffusion length.  
When $L_T \gg \xi'$, $f_1(\xi' / L_T) = A_0 + A_1\,\xi'/L_T$, 
which is equivalent to Eq.~(9).  
In the critical region, where  $L_T \ll \xi' \rightarrow \infty$,  
Eq.~(12) should be reduced to 
\begin{equation}
\sigma = C_1\,\frac{e^2}{\hbar L_T}\;.  
\end{equation}
Combining this equation and Eq.~(11), 
they obtained $\sigma \propto T^{1/3}$.  
The $T^{1/3}$ dependence was also predicted 
from numerical calculations that consider solely 
the effect of disorder~[22].

\begin{figure}
\begin{center}
\epsfig{file=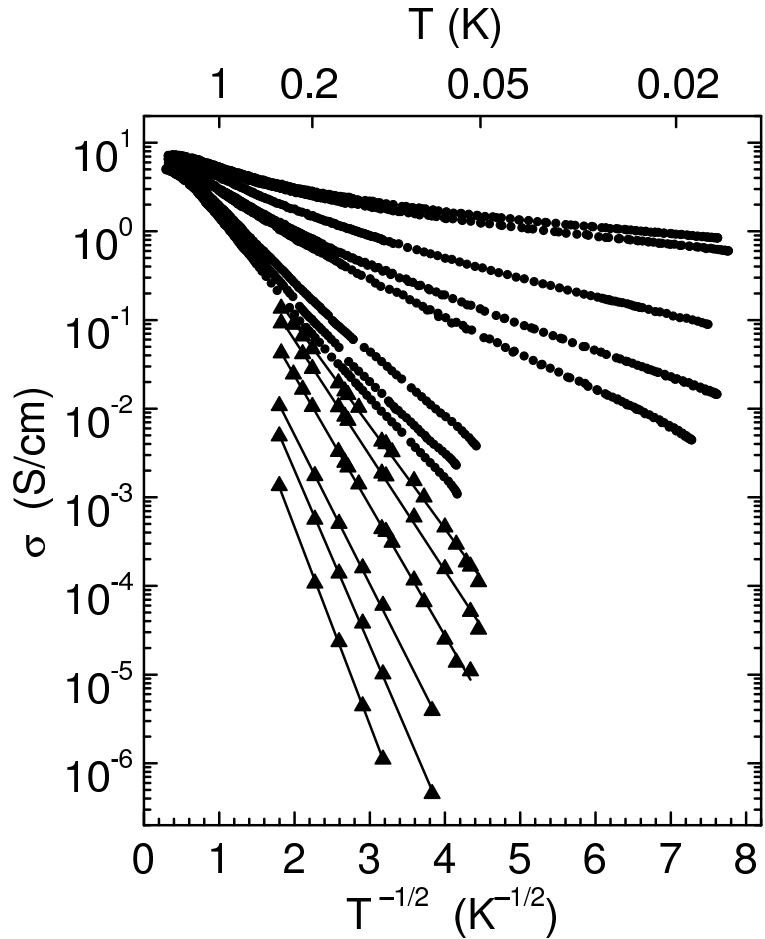,width=0.65\columnwidth,
bbllx=44,
bblly=134,
bburx=538,
bbury=744,clip=,
angle=0}
\end{center}
\caption
{
The logarithm of the conductivity as a function of $T^{-1/2}$ 
for insulating samples.  
the concentrations from bottom to top in units of
10$^{17}$~cm$^{-3}$ are 1.717, 1.752, 1.779, 1.796, 1.805, 
1.823, 1.840, 1.842, 1.843, 1.848, 1.850, 1.853, 1.856, 
and 1.858, respectively.  
}
\end{figure}
It is not clear whether the argument in Refs.~21 or 22 
is applicable to the present system or not, but there is an experimental 
fact that $\sqrt{T}$ dependence of conductivity changes to $T^{1/3}$ 
as $N_c$ is approached.  It is important that we find a consistent method 
that allows the determination of $\sigma(0)$ for both the cases.
For this purpose, we follow Al'tshuler and Aronov's 
manipulation~[21] of eliminating $m$ and $D$ in 
Eqs.~(9)--(11) and obtain 
\begin{equation}
\label{eq:new}
\sigma(T)= \sigma(0) + m'\sqrt{T/\sigma(T)}\;, 
\end{equation} 
where $m'=eA\sqrt{(\partial n / \partial \mu)}\,$, which is temperature 
independent.  In the limit of $\Delta\sigma(T) \ll \sigma(0) \approx
 \sigma(T)\,$, this equation gives the same value of $\sigma(0)$ as 
Eq.~(9) does.  When $\sigma(0) \ll \sigma(T)$, it yields 
a $T^{1/3}$ dependence for $\sigma(T)$.  
Thus, it is applicable to both $\sqrt{T}$ and $T^{1/3}$ dependent 
conductivity.  
From today's theoretical understanding of the problem, 
Eqs.~(9) and (14) 
are valid only for $L_T \gg \xi'$, and their applicability to 
the critical region is not clear, because the higher-order terms 
of the $\beta$ function~[23] which were once erroneously 
believed to be zero do not vanish~[24].  
Nevertheless, we expect Eq.~(14) to be a good expression 
for describing the temperature dependence of all metallic samples 
because it expresses both $\sqrt{T}$ and $T^{1/3}$ dependences 
as limiting forms.  
Then, based on Eq.~(14), we plot $\sigma(T)$ vs 
$\sqrt{T/\sigma(T)}$ for the four close to $N_c$ samples 
in Fig.~3(a).  
The data points align on straight lines, 
which supports the adequacy of Eq.~(14).
The zero-temperature conductivity $\sigma(0)$ is obtained 
by extrapolating to $T=0$.  
The curve on the top of Fig.~3(a) is for the sample with 
the lowest $N$ among the ones showing $\sqrt{T}$ dependence at low 
temperatures, i.e., this sample has the largest value of 
$\Delta\sigma(T)/\sigma(0)$ among $\sqrt{T}$ samples.  The value of 
$\sigma(0)$ obtained for this particular sample using Eq.~(14) 
differs only by 0.6\% from the value determined by the conventional 
extrapolation assuming Eq.~(9).  
This small difference is comparable to the variation arising from the 
choice of the temperature range in which the fitting is performed.  
Therefore, the extrapolation method proposed here 
is compatible with the conventional method based on the 
$\propto\sqrt{T}$ extrapolation.  

Based on this analysis the MIT is found 
to occur between the first and second samples from the bottom 
in Fig.~3(a), i.e., $1.858\!\times\!10^{17}\,{\rm cm}^{-3} < N_c 
< 1.861\!\times\!10^{17}\,{\rm cm}^{-3}$.  Thus, unlike the case for 
Si:P~[5,7], $N_c$ is fixed already 
within an accuracy of 0.16\% and the evaluation of the critical 
exponent $\mu$ will {\em not} be affected by the ambiguity 
in the determination of $N_c$.  
Figure~3(b) shows the zero-temperature conductivity 
$\sigma(0)$ as a function of $N/N_c-1$ on a double logarithmic scale. 
A fit of Eq.~(1) (dotted line) is excellent all 
the way down to $(N/N_c-1) = 4\times10^{-4}$.
The fitting parameters are $\mu = 0.50 \pm 0.04$,  
$N_c = (1.860\pm0.002)\times10^{17}$~cm$^{-3}$, and 
$\sigma^*=(40\pm2)$~S/cm. 
We note that $\mu = 0.46 \pm 0.18$ is obtained even when 
we use only the four samples closest to $N_c$ for the fitting.

\Head{Variable-range-hopping conduction in insulating samples 
and the critical exponents for localization length and dielectric susceptibility}
\begin{figure}
\begin{center}
\epsfig{file=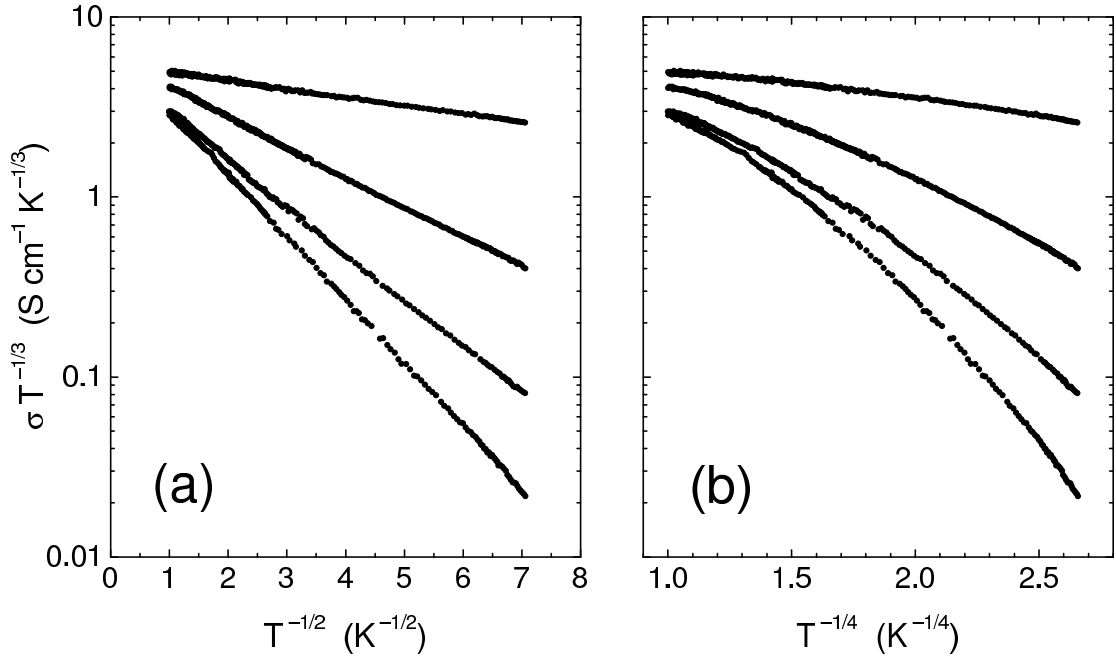,width=0.95\columnwidth,
bbllx=50,
bblly=376,
bburx=556,
bbury=674,clip=,
angle=0}
\end{center}
\caption
{
Conductivity multiplied by $T^{-1/3}$ 
vs (a) $T^{-1/2}$ and (b) $T^{-1/4}$.  
From bottom to top in units of 10$^{17}$~cm$^{-3}$, 
the concentrations are 1.848, 1.850, 1.853, and 1.856, 
respectively.  
}
\end{figure}
The temperature dependence of the conductivity $\sigma(T)$ 
of insulating samples is shown in Fig.~4.  
The electrical conduction of doped semiconductors 
on the insulating side of the MIT is often dominated 
by variable-range hopping (VRH) at low temperatures.  
The temperature dependence of $\sigma(T)$ 
for VRH is written in the form of 
\begin{equation}
\label{eq:VRH}
\sigma(T) = \tilde{\sigma}(T) \exp [-(T_0/T)^p\,],  
\end{equation}
where $p=1/2$ for the excitation within a parabolic-shaped energy gap 
(Coulomb gap), and $p=1/4$ for a constant single-particle density of states 
around the Fermi level~[13].  
The temperature dependence of $\tilde{\sigma}(T)$ 
contributes greatly to the temperature dependence of $\sigma(T)$ near $N_c$ 
because the factor $T_0/T$ in the exponential term becomes very small, i.e.,  
the temperature dependencies of the prefactor and that of the exponential term 
become comparable.  
Theoretically, $\tilde{\sigma}(T)$ is expected to depend on temperature as 
\begin{equation}
\tilde{\sigma}(T) \propto T^{\,r}
\end{equation}
but the value of $r$ including the sign has not been derived yet 
for VRH conduction with both $p=1/2$ and $p=1/4$. 

As we have seen in Fig.~2(b), the temperature 
variation of the low-temperature conductivity 
of the $^{70}$Ge:Ga samples within $\pm$0.3\% of $N_c$ is 
proportional to $T^{1/3}$.  
Since both the $T^{1/3}$ dependence of the conductivity and 
the VRH with $p=1/2$ are results of the electron-electron
interaction in disordered systems, 
they can be expressed, in principle, in a unified form.
Moreover, the electronic transport in barely metallic samples and 
that in barely insulating samples 
should be essentially the same at high temperatures so long as 
the inelastic scattering length and the thermal diffusion length 
are smaller than, or at most comparable to  
the correlation length or the localization length.  
So, the temperature dependence of conductivity at high temperatures 
should be the same on both sides of the transition. Such behavior is 
confirmed experimentally in the present system, 
i.e., as seen in Fig.~2(b) the conductivity of samples 
very close to $N_c$ shows a $T^{1/3}$ dependence at $T\approx0.5$~K, 
irrespective of the phase (metal or insulator) 
to which they belong at $T=0$.
Based on this consideration we fix $r=1/3$.
Figure~5 shows $\sigma T^{-r}$ with $r=1/3$ 
for four samples ($N/N_c=0.993$, 0.994, 0.996, and 0.998) 
as a function of (a) $T^{-1/2}$ and (b) $T^{-1/4}$.  
All the data points lie on straight lines with $p=1/2$ in 
Fig.~5(a) while they curve downward with $p=1/4$ 
in Fig.~5(b).  
This dependence is maintained even when we change the values of $r$ 
between $r=1/2$ and 1/4.  Thus we conclude that the 
conductivity of all samples on the insulating side for $N$ up to 
0.998$N_c$ is described by the theory for the VRH conduction 
where the excitation occurs within the Coulomb gap, 
i.e., Eq.~(15) with $p=1/2$.

Based on these findings, we evaluate the $N$ dependence of $T_0$ in 
Eq.~(15) with $p=1/2$ and $r=1/3$.  
Figure~6 shows $T_0$ as a function of $1-N/N_c$.  
\begin{figure}
\begin{center}
\epsfig{file=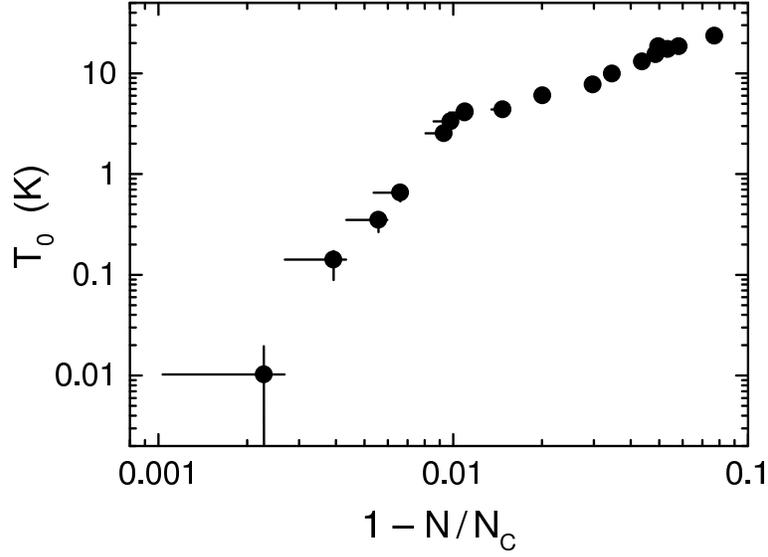,width=0.65\columnwidth,
bbllx=28,
bblly=332,
bburx=557,
bbury=724,clip=,
angle=0}
\end{center}
\caption
{
$T_0$ determined by $\sigma(T) \propto 
T^{1/3}\exp[-(T_0/T)^{1/2}]$ as a function of 
the dimensionless concentration $1-N/N_c$.
}
\end{figure}
The vertical and horizontal error bars have been estimated 
based on the values of $T_0$ obtained with $r=1/2$ and $r=1/4$, 
and the values of $1-N/(1.858\times10^{17}$~cm$^{-3})$ and $1-N/ 
(1.861\times10^{17}$~cm$^{-3})$, where $1.858\times10^{17}$~cm$^{-3}$ 
is the highest concentration in the insulating phase and 
$1.861\times10^{17}$~cm$^{-3}$ is the lowest in the metallic phase, 
respectively.  
According to theory~[13], $T_0$ in Eq.~(15) 
is given by 
\begin{equation}
\label{eq:T0}
k_B T_0 \approx \frac{1}{\,4\pi\epsilon_0\,}\,
\frac{2.8 e^2}{\,\epsilon (N)\, \xi (N)\,}
\end{equation}
in SI units, where $\epsilon (N)$ 
is the dielectric constant, and $\xi (N)$ is the localization length. 
Here, we should note that the condition $T<T_0$ is needed 
for the VRH theory to be valid, 
i.e., $T_0$ has to be evaluated only from the data 
obtained at temperatures low enough to satisfy the above condition.  
This requirement is fulfilled in Fig.~6 
for all the samples except for the one with $N=0.998N_c$.  
Concerning this latter sample, we will include it for the determination 
of $\gamma$ (Fig.~8) and $\xi$ and $\chi_{\rm imp}$ 
(Fig.~9) but not for the calculation of the critical exponents.  

Our next step is to separate $T_0$ into $\epsilon$ and $\xi$ 
in the framework of the theory of VRH conduction 
with and without a weak magnetic field~[13].  
For $\xi/\lambda\ll1$, the magnetoconductance is expressed as 
\begin{equation}
\label{eq:rhoinB}
-\ln\left[\frac{\,\sigma(B,T)\,}{\sigma(0,T)}\right] 
\approx 0.0015\left(\frac{\,\xi\,}{\,\lambda\,}\right)^{\!4}
\left(\frac{\,T_0\,}{T}\right)^{3/2},
\end{equation}
where $\lambda\equiv\sqrt{\hbar/eB}$ 
is the magnetic length in SI units. 
According to Eq.~(18), 
the magnetic-field variation of $\ln\sigma$ at $T={\rm const.}$ 
is proportional to $B^2$, i.e., 
\begin{equation}
-\ln\sigma(B,T) = -\ln\sigma(0,T) + C_2 (T)B^2, 
\end{equation}
and the slope $C_2 (T)$ in the above equation is proportional 
to $T^{-3/2}$.  In order to demonstrate that these relations hold 
for our samples, we show for the $N=0.989N_c$ 
sample $-\ln\sigma(B,T)$ vs $B^2$ in Fig.~7(a) 
and $C_2 (T)$ determined by least-square fitting of 
$-\partial\ln\sigma/\partial B^2$ 
vs $T^{-3/2}$ in Fig.~7(b).  
\begin{figure}
\begin{center}
\epsfig{file=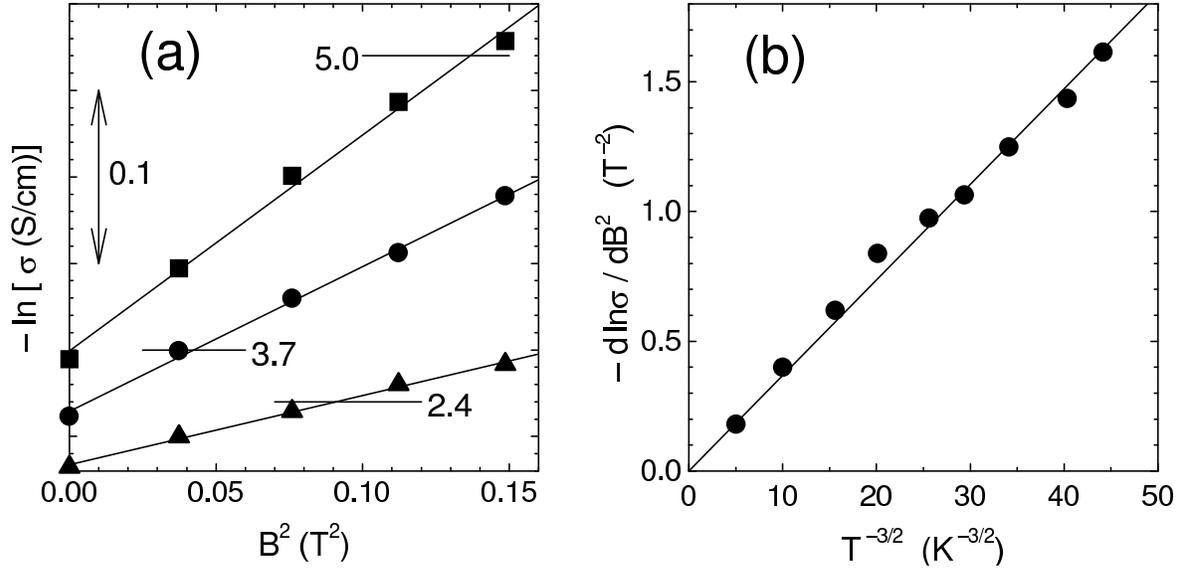,width=\columnwidth,
bbllx=29,
bblly=378,
bburx=563,
bbury=637,clip=,
angle=0}
\end{center}
\caption
{
(a) Logarithm of $\sigma^{-1}$ vs $B^2$ at constant 
temperatures for the sample having $N=1.840\times10^{17}$~cm$^{-3}$.   
From top to bottom the temperatures are 0.095~K, 0.135~K, and 0.215~K, 
respectively.  The solid lines represent the best fits.  
(b)~Slope $-d\ln\sigma/dB^2$ vs $T^{-3/2}$ 
for the same sample.  
The solid line represents the best fit.  
}
\end{figure}
Since Eq.~(18) is equivalent to 
\begin{equation} 
\label{eq:gamma}
\gamma\equiv C_2 (T)/T^{-3/2}\approx
0.0015\,(e/\hbar)^2\,\xi^{\,4}\,T_0^{\,3/2}\,, 
\end{equation} 
$\xi$ is given by
\begin{equation} 
\label{eq:xi}
\xi \approx 5.1\,(\hbar/e)^{1/2}\,\gamma^{1/4}\,T_0^{\,-3/8}\,.  
\end{equation}
In this way we have determined $\gamma$ as a function of $T_0$ for nine samples 
(Fig.~8).  
\begin{figure}
\begin{center}
\epsfig{file=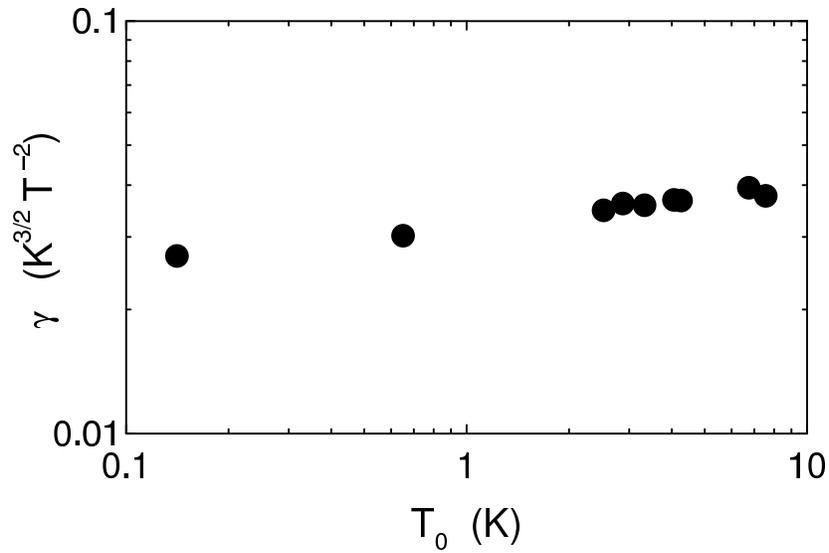,width=0.7\columnwidth,
bbllx=68,
bblly=429,
bburx=530,
bbury=739,clip=,
angle=0}
\end{center}
\caption
{
Coefficient $\gamma$ defined by Eq.~(20) 
as a function of $T_0$.  
}
\end{figure}
The value of $\gamma$ is almost independent of $T_0$, 
and if one assumes the form of $\gamma \propto T_0^{\,\delta}$, 
one obtains a small value of $\delta = 0.094 \pm 0.005$ 
from least-square fitting.  
We determine $\xi$ and $\chi_{\rm imp}=\epsilon-\epsilon_h$ 
from Eqs.~(17) and (21), 
and show them in Fig.~9 
as a function of $1-N/N_c$ and $N_c/N-1$, respectively. 
Here, $\epsilon_h$ is the dielectric constant of the host Ge, 
and hence, $\chi_{\rm imp}$ is the dielectric susceptibility 
of the Ga acceptors.  
We should note that both $\xi$ and $\chi_{\rm imp}$ are sufficiently larger 
than the Bohr radius (8~nm for Ge) and $\epsilon_h=15.4$~[25], respectively. 
According to the theories of the MIT, both $\xi$ and $\chi_{\rm imp}$ 
diverge at $N_c$ as $\xi (N) \propto (1-N/N_c)^{-\nu}$
and $\chi_{\rm imp}(N)\propto(N_c/N-1)^{-\zeta}$, respectively.  
We find, however, both $\xi$ and $\chi_{\rm imp}$ do not show such 
simple dependencies on $N$ in the range shown in Fig.~9, 
and that there is a sharp change of both dependencies at $N\approx0.99N_c$.  
\begin{figure}
\begin{center}
\epsfig{file=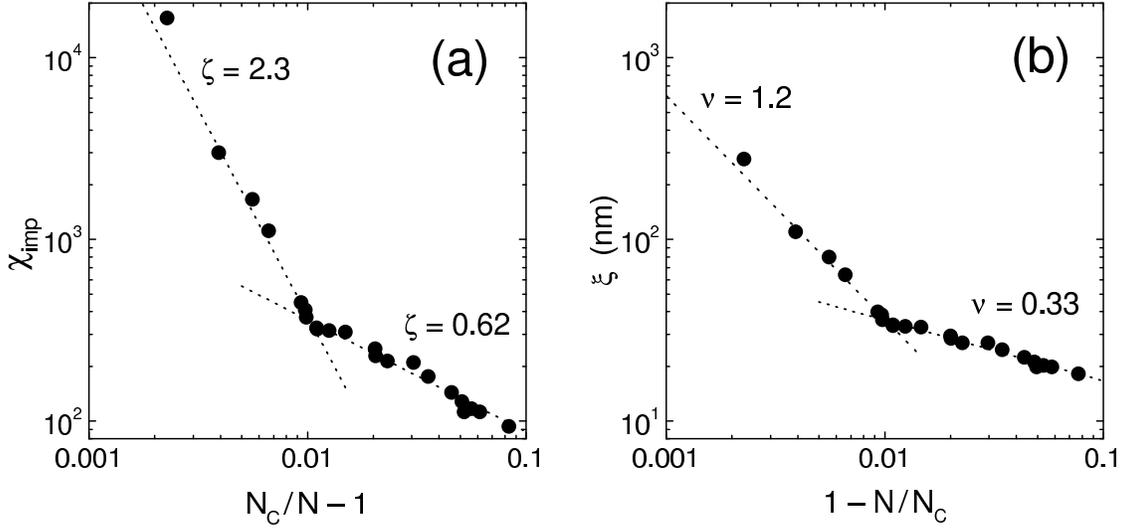,width=0.95\columnwidth,
bbllx=29,
bblly=376,
bburx=573,
bbury=638,clip=,
angle=0}
\end{center}
\caption
{
(a) Dielectric susceptibility $\chi_{\rm imp}$ arising 
from the impurities vs $N_c/N-1$.  
(b)~ Localization length $\xi$ vs $1-N/N_c$.  
}
\end{figure}
On both sides of the change in slope, the concentration dependence 
of $\xi$ and $\chi_{\rm imp}$ are expressed well by the scaling formula 
as shown in Fig.~9.  
Theoretically, the quantities should show the critical behavior 
when $N$ is very close to $N_c$.  So $\nu=1.2\pm0.3$ and $\zeta=2.3\pm0.6$ 
may be concluded from the data in $0.99<N/N_c$.  
However, the other region ($0.9<N/N_c<0.99$), where 
we obtain $\nu=0.33\pm0.03$ and $\zeta=0.62\pm0.05$, 
is also very close to $N_c$ in a conventional experimental sense.  

As a possible origin for the change in slope, we refer to the 
effect of compensation.  Although our samples are nominally 
uncompensated, doping compensation of less than 0.1\% 
may be present due to residual isotopes that become {\it n}-type impurities 
after NTD.  In addition to the doping compensation, the effect known 
as ``self compensation" may play an important role near $N_c$~[26]. 
It is empirically known that the doping compensation 
affects the value of the critical exponents~[11]. 
Rentzsch {\em et al.} studied VRH conduction 
of {\em n}-type NTD Ge in the concentration range of $0.2<N/N_c<0.91$, 
and showed that $T_0$ vanishes as $T_0\propto (1-N/N_c)^{\alpha}$ with 
$\alpha \approx 3$ for $K=38$\% and 54\%, where $K$ is the compensation 
ratio~[27].  
Since $\alpha \approx \nu + \zeta$ [Eq.~(17)], we find for 
our NTD $^{70}$Ge:Ga samples 
$\alpha=3.5\pm0.8$ for $0.99<N/N_c<1$ and $\alpha=0.95\pm0.08$ 
for $0.9<N/N_c<0.99$.  Interestingly, $\alpha = 3.5\pm0.8$ agrees 
with $\alpha \approx 3$ found for compensated samples.  
Moreover, we have recently proposed the possibility that the conductivity 
critical exponent $\mu \approx 1$ in the same 
$^{70}$Ge:Ga only within the very vicinity of $N_c$ 
(up to about +0.1\% of $N_c$)~[28]. 
An exponent of $\mu=0.50\pm0.04$, on the other hand, holds for a 
wider region of $N$ up to $1.4N_c$ as we have seen in Fig.~3(b).  
Again, $\mu \approx 1$ near $N_c$ may be viewed as the 
effect of compensation.  Therefore, it may be possible that the region of $N$ 
around $N_c$ where $\nu\approx1$ and $\mu\approx1$ 
changes its width as a function of the doping compensation.  
In the limit of zero compensation, 
the part which is characterized by 
$\nu\approx1$ and $\mu\approx1$ vanishes, i.e., 
we propose $\nu=0.33\pm0.03$, $\zeta=0.62\pm0.05$, and 
$\mu=0.50\pm0.04$ for truly uncompensated systems and that 
the relation $\mu=\nu$~[3] is not satisfied. 
In compensated systems, on the other hand, $\mu=\nu$ may hold 
as it does in the very vicinity of $N_c$.  
However, the preceding discussion needs to be proven 
experimentally in the future by using samples whose 
compensation ratios are controlled precisely 
and systematically.

\Head{Metal-insulator transition in magnetic fields}

Figure~10 shows the temperature dependence of the conductivity 
of the sample having $N=2.004\times10^{17}$~cm$^{-3}$ 
for several values of the magnetic induction $B$.  
\begin{figure}
\begin{center}
\epsfig{file=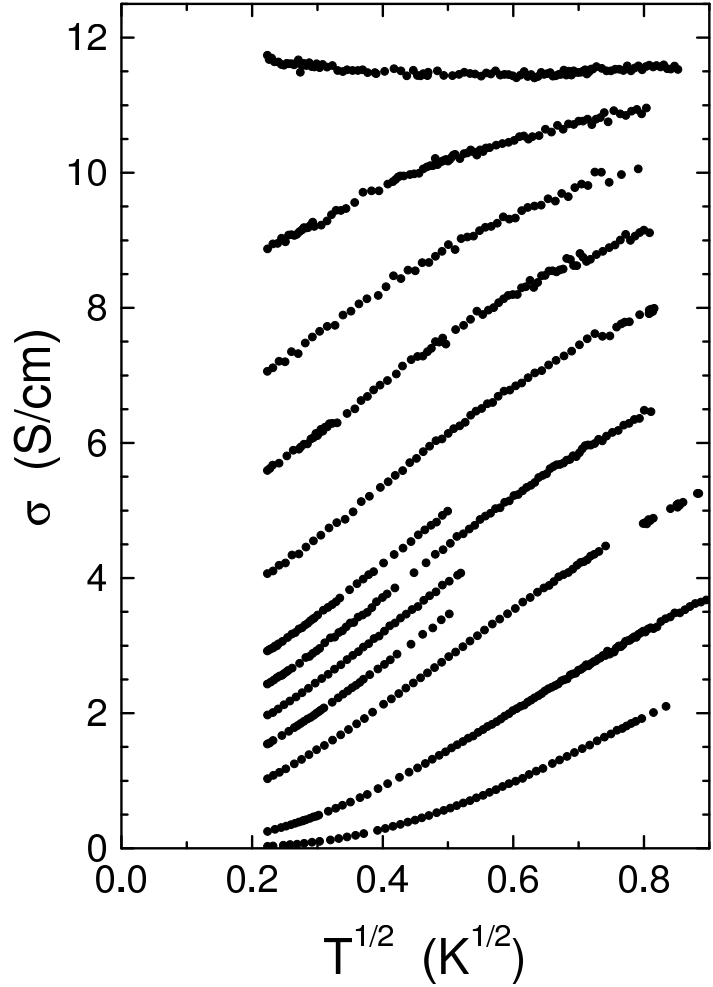,width=0.6\columnwidth,
bbllx=44,
bblly=100,
bburx=521,
bbury=775,clip=,
angle=0}
\end{center}
\caption
{
Conductivity of the sample having $N=2.004\times10^{17}$~cm$^{-3}$ 
as a function of $T^{1/2}$ at several magnetic fields.  
The values of the magnetic induction 
from top to bottom in units of tesla are 0.0, 1.0, 2.0, 3.0, 4.0, 
4.7, 5.0, 5.3, 5.6, 6.0, 7.0, and 8.0, respectively.
}
\end{figure}
Application of the magnetic field decreases the conductivity 
and eventually drives the sample into the insulating phase.  
This property can be understood in terms of the shrinkage 
of the wave function due to the magnetic field.  
In strong magnetic fields and at low temperatures, i.e., 
when $g \mu_B B \gg k_B T$, the conductivity shows another 
$\sqrt{T}$ dependence~[1]
\begin{equation}
\label{eq:eeinB1}
\sigma(N,B,T) = \sigma(N,B,0) + m_B(N,B)\,\sqrt{T}\,,
\end{equation}
where 
\begin{equation}
\label{eq:mB}
m_B \,=\, \frac{\;e^2}{\hbar}\frac{1}{\;4\pi^2\;}\frac{1.3}{\;\sqrt{2}\;}
\left(\frac{4}{\;3\;}-\frac{1}{\;2\;}\tilde{F} \right) 
\sqrt{\frac{k_B}{\;\hbar D\;}}\;.
\end{equation}
One should note that Eqs.~(9) and (22) are valid 
only in the limits of $[\sigma (N,0,T) - \sigma (N,0,0)] \ll \sigma(N,0,0)$ 
or $[\sigma(N,B,T) - \sigma(N,B,0)] \ll \sigma(N,B,0)$. 
It is for this reason that we have observed a $T^{1/3}$ dependence 
rather than the $\sqrt{T}$ dependence at $B=0$ in Fig.~2 
as the critical point [$\sigma(N,0,0)=0$] is approached 
from the metallic side.  
However, Fig.~10 shows that the $\sqrt{T}$ dependence 
holds when $B\neq0$ even around the critical point. 
Hence, we use Eq.~(22) to evaluate the zero-temperature 
conductivity $\sigma(N,B,0)$ in magnetic fields.  
Since $m_B$ is independent of $B$, the conductivity for various values 
of $B$ plotted against $\sqrt{T}$ should appear as a group of parallel lines.  
This is approximately the case as seen in Fig.~10 at low 
temperatures (e.g., $T<0.25$~K).  

The zero-temperature conductivity $\sigma(N,B,0)$ 
in various magnetic fields obtained by extrapolation of 
$\sigma(N,B,T)$ to $T=0$ based on Eq.~(22) is shown 
in Fig.~11.
\begin{figure}
\begin{center}
\epsfig{file=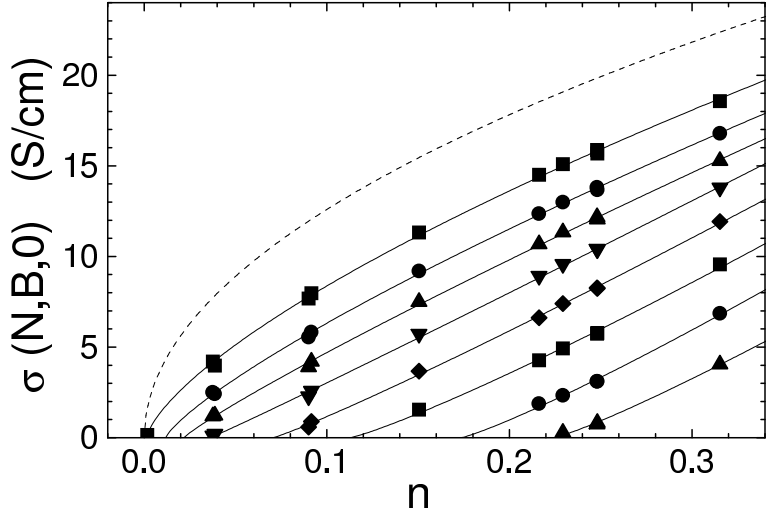,width=0.65\columnwidth,
bbllx=24,
bblly=405,
bburx=566,
bbury=771,clip=,
angle=0}
\end{center}
\caption
{
Zero-temperature conductivity $\sigma(N,B,0)$ vs normalized 
concentration $n \equiv [\sigma(N,0,0)/\sigma^*(0)]^{2.0} 
= N/N_c(0)-1$, where $\sigma(N,0,0)$ is the zero-temperature 
conductivity and $\sigma^*(0)$ is the prefactor both at $B=0$. 
From top to bottom the magnetic induction increases from 1~T to 8~T 
in steps of 1~T.  The dashed curve at the top is for $B=0$.  
The solid curves represent fits of $\sigma(N,B,0) \propto 
[n/n_c(B)-1]^{\mu(B)}$.  For $B\geq6$~T, we assume $\mu=1.15$.  
}
\end{figure}
Here, $\sigma(N,B,0)$ is plotted as a function of the normalized concentration: 
\begin{equation}
\label{eq:n}
n \equiv [\sigma(N,0,0)/\sigma^*(0)]^{2.0}.  
\end{equation}
Since the relation between $N$ and $\sigma(N,0,0)$ was 
established in Fig.~3(b) as
$\sigma(N,0,0) = \sigma^*(0)[N/N_c(0)-1]^{0.50}$ 
where $N_c(0)=1.860\!\times\!10^{17}\,{\rm cm}^{-3}$ 
and $\sigma^*(0)=40$~S/cm, 
$n$ is equivalent to $N/N_c(0)-1$.  
Henceforth, we will use $n$ instead of $N$ because employing $n$ 
reduces the scattering of the data caused by several experimental 
uncertainties, and it further helps us concentrate on observing how 
$\sigma(N,B,0)$ varies as $B$ is increased.  
Similar evaluations of the concentration have been used by various groups.  
In their approach, the ratio of the resistance at 4.2~K to that at a room 
temperature is used to determine the concentration~[5].  
The dashed curve in Fig.~11 is for $B=0$, which 
merely expresses Eq.~(24), 
and the solid curves represent fits of 
\begin{equation}
\label{eq:nMIT}
\sigma(N,B,0) = \sigma_0(B)[n/n_c(B)-1]^{\mu(B)}.  
\end{equation}
The exponent $\mu(B)$ increases from 0.5 with increasing $B$ 
and reaches a value close to unity at $B\geq4$~T.  For example, 
$\mu=1.03\pm0.03$ at $B=4$~T and $\mu=1.09\pm0.05$ at $B=5$~T.  
When $B\geq6$~T, three-parameter [$\sigma_0(B)$, 
$n_c(B)$, and $\mu(B)$] fits no longer give reasonable results  
because the number of samples available for the fit 
decreases with increasing $B$.  Hence, we give the solid curves 
for $B\geq6$~T assuming $\mu(B)=1.15$.  

We show $\sigma(N,B,0)$ as a function of $B$ 
in Fig.~12 for three different samples.  
\begin{figure}
\begin{center}
\epsfig{file=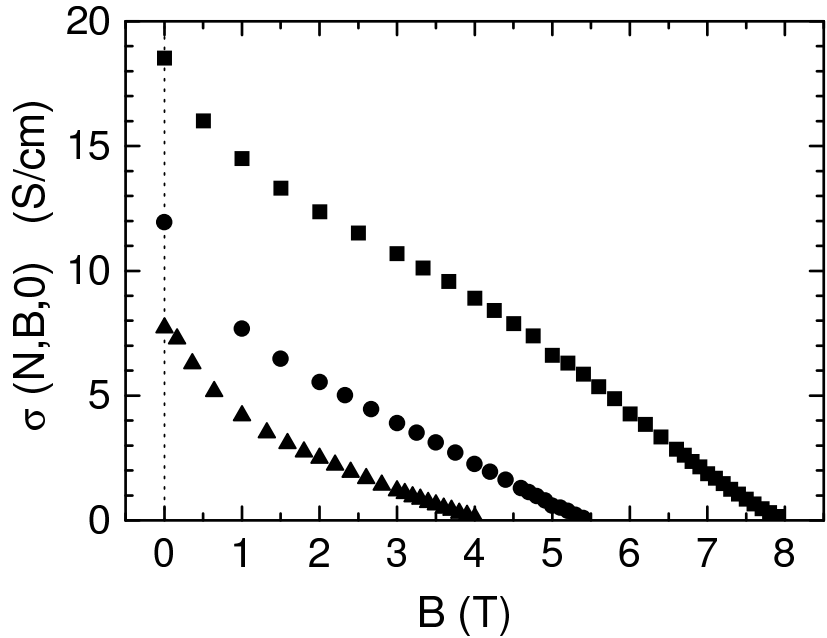,width=0.7\columnwidth,
bbllx=56,
bblly=355,
bburx=550,
bbury=740,clip=,
angle=0}
\end{center}
\caption
{
Zero-temperature conductivity $\sigma(N,B,0)$ 
vs magnetic induction $B$.  From bottom to top, 
the normalized concentrations defined by Eq.~(24) 
are 0.04, 0.09, and 0.22, respectively.  
}
\end{figure}
When the magnetic field is weak, i.e., the correction 
$\Delta\sigma_B(N,B,0)\equiv\sigma(N,B,0)-\sigma(N,0,0)$ 
due to $B$ is small compared with $\sigma(N,0,0)$, the field 
dependence of $\Delta\sigma_B(N,B,0)$ looks consistent with 
the prediction by the interaction theory~[1], 
\begin{equation}
\label{eq:sB0}
\Delta\sigma_B(N,B,0)\;=\;-\frac{\;e^2}{\hbar}
\frac{\tilde{F}}{\;4\pi^2\;}
\sqrt{\frac{\;g\mu_BB\;}{2\hbar D}}\;\propto\sqrt{B}\,.
\end{equation} 
In larger magnetic fields, $\sigma(N,B,0)$ deviates 
from Eq.~(26) and eventually vanishes 
at some magnetic induction $B_c$.  
For the samples in Fig.~12, we tuned the magnetic 
induction to the MIT in a resolution of 0.1~T.  
\begin{figure}
\begin{center}
\epsfig{file=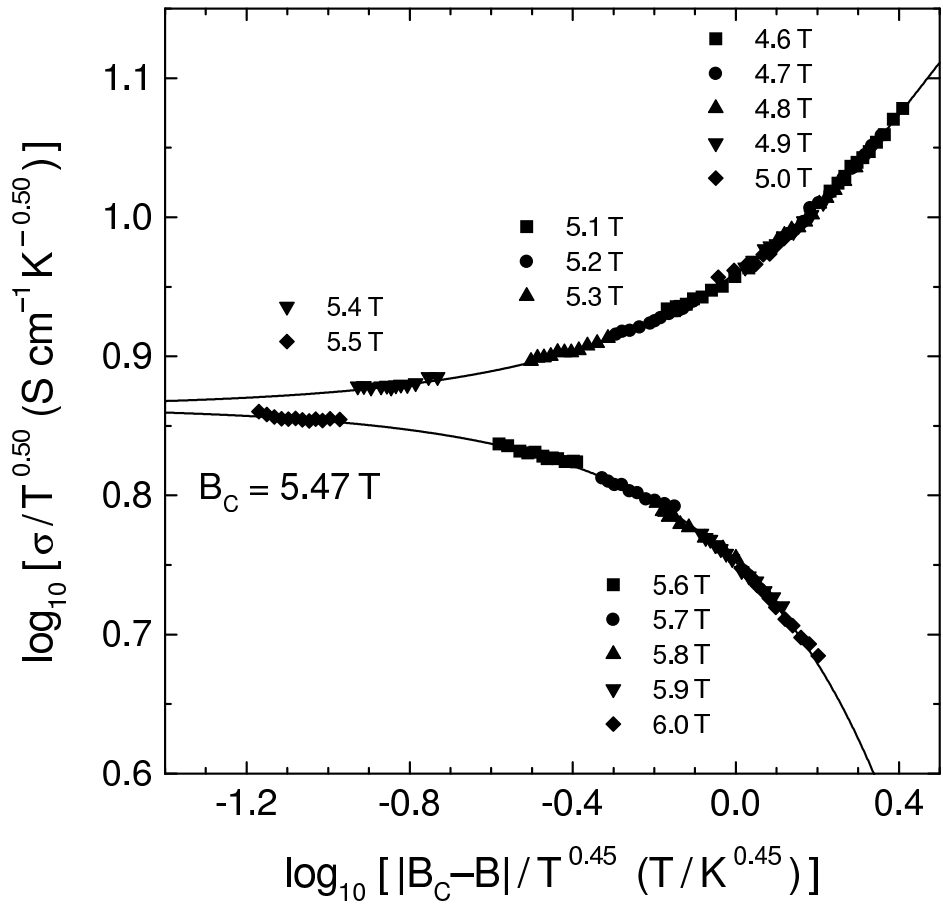,width=0.8\columnwidth,
bbllx=37,
bblly=283,
bburx=507,
bbury=737,clip=,
angle=0}
\end{center}
\caption
{
Finite-temperature scaling plot for 
the field-induced metal-insulator transition in 
the $^{70}$Ge:Ga sample having $n=0.09$.  
}
\end{figure}
We fit an equation similar to Eq.~(25),
\begin{equation}
\label{eq:BMIT}
\sigma(N,B,0) 
= \sigma_{0}'(n)[1-B/B_c(n)]^{\mu'(n)},  
\end{equation}
to the data close to the critical point.  
As a result we obtain $\mu'=1.1\pm0.1$ for all of the three samples.  
The value of $\mu'$ depends on the choice of the magnetic-field range 
to be used for the fitting, and this fact leads to the error of $\pm0.1$ 
in the determination of $\mu'$.  
In Fig.~13 we show that $\mu'=1.1$ yields an excellent 
finite-temperature scaling [Eq.~(3)].  
Note that the data both on the metallic side and on the insulating 
side are included in this scaling plot.  
Here we employ $B_c$ obtained by fitting Eq.~(27), 
$x=1/2$ from the fact that $\sqrt{T}$ dependence holds in magnetic 
fields even around the critical point, and $y=x/\mu'=0.45$, 
i.e., none of them are treated as a fitting parameter.  
Hence, Fig.~13 strongly supports $\mu'=1.1$.  

From the critical points $n_c(B)$ and $B_c(n)$, the phase diagram at $T=0$
is constructed on the $(N,B)$~plane as shown in Fig.~14. 
\begin{figure}
\begin{center}
\epsfig{file=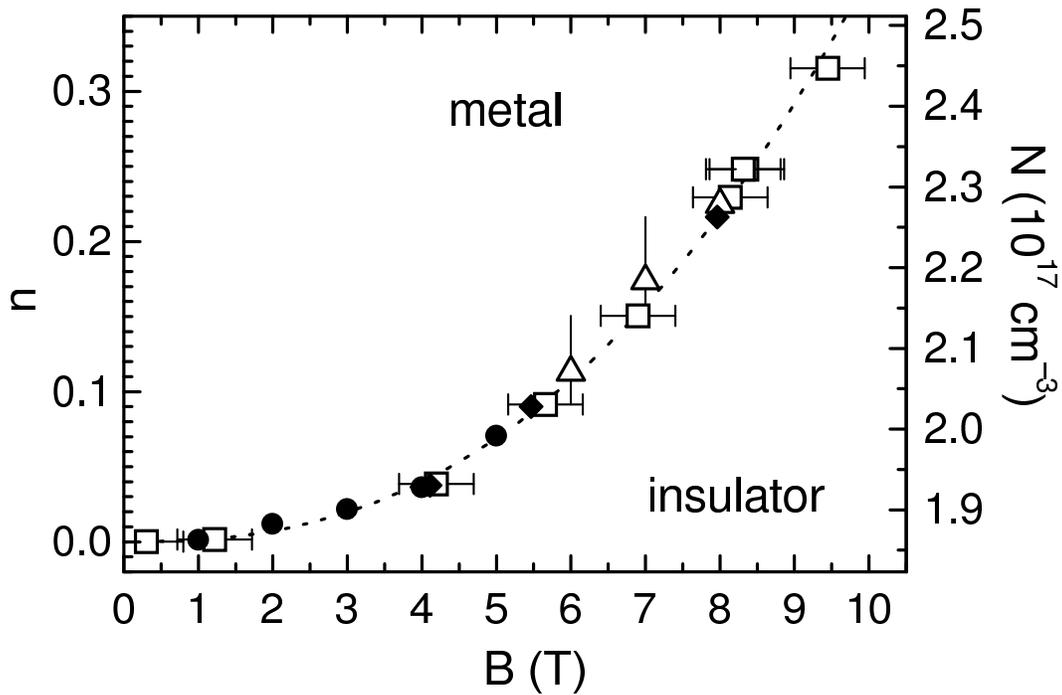,width=0.9\columnwidth,
bbllx=22,
bblly=370,
bburx=562,
bbury=725,clip=,
angle=0}
\end{center}
\caption
{
Phase diagram of $^{70}$Ge:Ga at $T=0$.  
The solid circles and the open triangles represent the critical 
concentrations $n_c$, and the solid diamonds and the open boxes 
represent the critical magnetic induction $B_c$.
}
\end{figure}
Here, $n_c(B)$ for $B\geq6$~T shown by triangles 
are obtained by assuming $\mu=1.15$.  The vertical solid lines associated 
with the triangles represent the range of values over which $n_c(B)$ have 
to exist, i.e., between the highest $n$ in the insulating phase 
and the lowest $n$ in the metallic phase.  Solid diamonds represent 
$B_c$ for the three samples in which we have studied 
the magnetic-field-induced MIT.  
Estimations of $B_c$ for the other samples are also shown 
by open boxes with error bars.  
The boundary between metallic phase and insulating phase is expressed 
by a power-law relation: 
\begin{equation}
\label{eq:boundary}
n = C_3\,B^\beta .
\end{equation}
From the eight data points denoted by the solid symbols, 
we obtain $C_3=(1.33 \pm 0.17)\times 10^{-3}$~T$^{-\beta}$ 
and $\beta=2.45\pm0.09$ as shown by the dotted curve.  
The shift of $N_c$ in magnetic fields was studied 
theoretically by Khmel'nitskii and Larkin~[29].  
They considered a noninteracting electron system starting from 
\begin{equation}
\label{eq:KL}
\sigma(N,B,0)\approx \frac{\,e^2}{\hbar\,\xi'}\:f_2(B^{\alpha'}\xi'), 
\end{equation}
where $\xi'$ is the correlation length.  They claimed that the argument of 
the function $f_2$ should be a power of the magnetic flux through a region 
with dimension $\xi'$.  This means 
\begin{equation}
\label{eq:KL2}
\sigma(N,B,0)\approx \frac{\,e^2}{\hbar\,\xi'}\:f_2(\xi'/\lambda), 
\end{equation}
where $\lambda\equiv\sqrt{\hbar/eB}$ is the magnetic length,  
and hence, $\alpha'=1/2$.  
In order to discuss the shift of the MIT due to the magnetic field,
they rewrote Eq.~(30) as 
\begin{equation}
\label{eq:KL5}
\sigma(N,B,0)\approx \frac{\,e^2}{\hbar\,\lambda}\:
                     \phi (t \, \lambda^{1/\nu}),
\end{equation}
based on the relation in zero magnetic field
\begin{equation}
\xi'\propto t^{-\nu}.
\end{equation}
Here, $t$ is a measure of distance from the critical point 
in zero field, e.g., 
\begin{equation}
t\equiv[N/N_c(0)-1].
\end{equation}
The zero point of the function $\phi$ gives the MIT, and 
the shift of the critical point for the MIT equals  
\begin{equation}
N_c(B)-N_c(0) \propto B^{1/2\nu}.
\end{equation}
Thus, $\beta = 1/(2\nu)$ results.  
In the present system, however, 
this relation does not hold, as long as we assume $\mu=\nu$~[3].  
Experimentally, we find $\beta=2.5$, while $1/(2\nu)=1/(2\mu)=1$
for $^{70}$Ge:Ga at $B=0$.  

Based on the phase diagram we shall consider the relationship between 
the two critical exponents: $\mu$ for the doping-induced MIT and 
$\mu'$ for the magnetic-field-induced MIT.  
Suppose that a sample with normalized concentration $n$ 
has a zero-temperature conductivity $\sigma$ at $B\neq0$ 
and that $[n/n_c(B)-1]\ll 1$ or $[1-B/B_c(n)]\ll 1$.  
From Eqs.~(25) and (27), 
we have two expressions for $\sigma$:  
\begin{equation}
\label{eq:n1MIT}
\sigma = \sigma_0 \,(n/n_c-1)^\mu
\end{equation}
and
\begin{equation}
\label{eq:B1MIT}
\sigma = \sigma_0'\,(1-B/B_c)^{\mu'}.
\end{equation}
On the other hand, we have from Eq.~(28) 
\begin{equation}
\label{eq:approx1}
n/n_c = (B/B_c)^{-\beta} = [1-(1-B/B_c)]^{-\beta}
\approx1+\beta(1-B/B_c)
\end{equation}
in the limit of $(1-B/B_c)\ll1 $. 
This equation can be rewritten as 
\begin{equation}
\label{eq:approx2}
 (n/n_c-1)/\beta \,\approx\, (1-B/B_c). 
\end{equation}
Using Eqs.~(35), (36), and (38), we obtain 
\begin{equation}
\label{eq:always}
\sigma_0'(1-B/B_c)^{\mu'} \approx \beta^\mu\sigma_0 (1-B/B_c)^\mu. 
\end{equation}
Since Eq.~(39) has to hold for arbitrary $B$, 
the following relations 
\begin{equation}
\label{eq:s0's0}
\sigma_0'=\beta^\mu\sigma_0
\end{equation}
and 
\begin{equation}
\label{eq:mu'mu}
\mu'=\mu
\end{equation}
are derived. 

\begin{figure}
\begin{center}
\epsfig{file=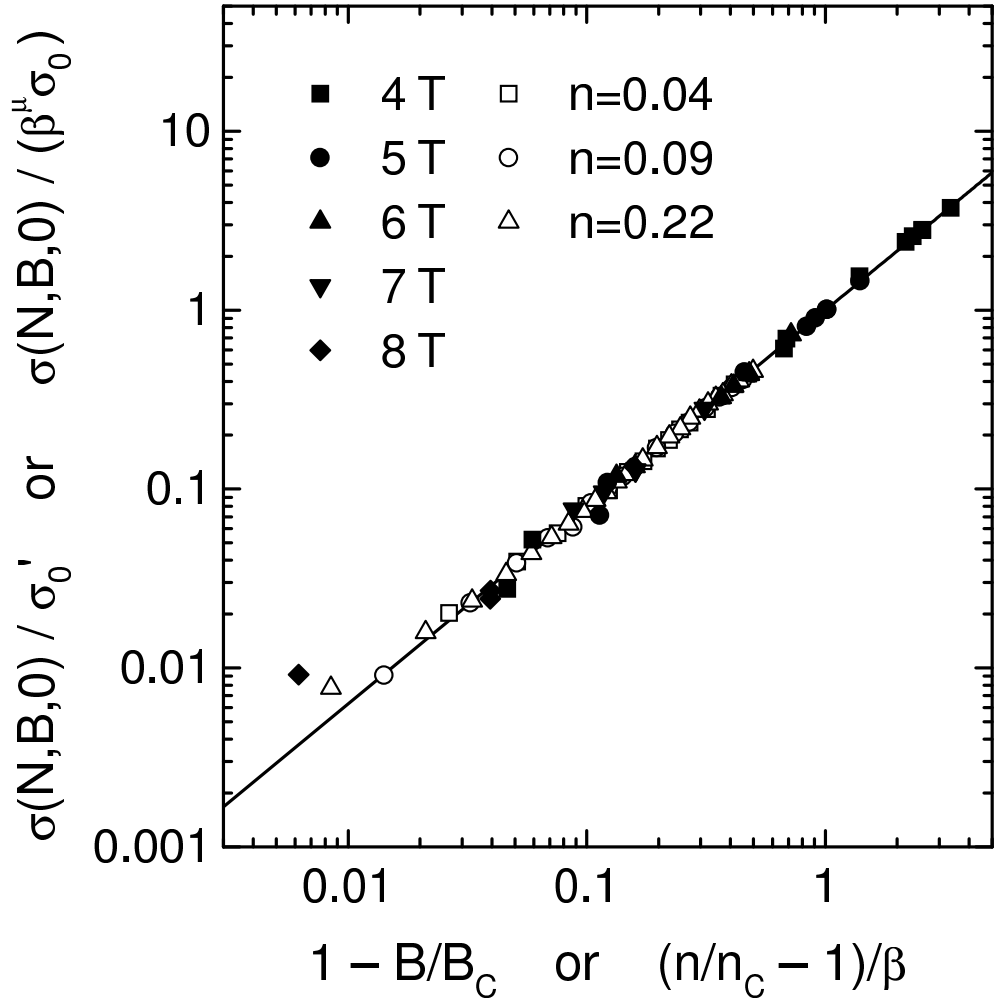,width=0.85\columnwidth,
bbllx=37,
bblly=218,
bburx=543,
bbury=731,clip=,
angle=0}
\end{center}
\caption
{
Normalized zero-temperature conductivity $\sigma(N,B,0)
/\sigma_0'(n)$ and $\sigma(N,B,0)/[\beta^\mu\,\sigma_0(B)]$ 
as functions of $[1-B/B_c(n)]$ and $[n/n_c(B)-1]
/\beta$, respectively, where $\beta=2.5$ and $\mu=1.1$. 
The solid line denotes a power-law behavior with the exponent of 1.1.  
The open and solid symbols represent the results of 
the magnetic-field-induced metal-insulator transition (MIT) 
in the range $(1-B/B_c)<0.5$ for three different samples 
($n=0.04$, 0.09, and 0.22) and the doping-induced MIT 
in constant magnetic fields (4, 5, 6, 7, and 8~T), respectively.
}
\end{figure}
In Fig.~15 we see how well Eq.~(41) holds for 
the present system.  
We have already shown in Fig.~12
that $\mu'= 1.1 \pm 0.1$ is practically independent of $n$.
Concerning the exponent $\mu$, however, its dependence on $B$ has not 
been ruled out completely even for the highest $B$ we used 
in the experiments.  This is mainly because the number of available data points 
at large $B$ is not sufficient for a precise determination of $\mu$.  
In Fig.~15 the results of the doping-induced MIT for $B\geq4$~T 
(solid symbols) and the magnetic-field-induced MIT 
for three different samples (open symbols) are plotted.  
Here, we plot $\sigma(N,B,0)/[\beta^\mu\,\sigma_0(B)]$ vs 
$[n/n_c(B)-1]/\beta$ with $\beta=2.5$ and $\mu=1.1$ 
for the doping-induced MIT, and $\sigma(N,B,0)/\sigma_0(B)'$ 
vs $[1-B/B_c(n)]$ for the magnetic-field-induced MIT.  
Figure~15 shows that the data points align exceptionally well 
along a single line describing a single exponent $\mu=\mu'=1.1$.  

We saw in Fig.~11 that $\mu$ apparently takes 
smaller values in $B\leq3$~T, which seemingly contradicts 
the above consideration. We can understand this as follows.  
We find that the critical exponent $\mu$ in zero magnetic field 
is 0.5 which is different from the values of $\mu$ in magnetic fields.
Hence, one should note whether the system under consideration 
belongs to the ``magnetic-field regime" or not.
In systems where the MIT occurs, there are several characteristic 
length scales: the correlation length, the thermal diffusion length, 
the inelastic scattering length, the spin scattering length,
the spin-orbit scattering length, etc.
As for the magnetic field, it is characterized by the magnetic length 
$\lambda\equiv\sqrt{\hbar/eB}$.  When $\lambda$ is smaller than 
the other length scales, the system is in the ``magnetic-field regime."
As the correlation length $\xi'$ diverges at the MIT, $\lambda < \xi'$ holds 
near the critical point, no matter how weak the magnetic field is. 
When the field is not sufficiently large, the ``magnetic-field regime" 
where we assume $\mu = 1.1$ to hold, is restricted to a narrow region of 
concentration.  Outside the region, the system crosses over to 
the ``zero-field regime" where $\mu = 0.5$ is expected.  
This is what is seen in Fig.~11.

\HEAD{Conclusion}

We have measured the electrical conductivity of NTD $^{70}$Ge:Ga 
to study the metal-insulator transition, ruling out an ambiguity due to 
inhomogeneous distribution of impurities.  
The critical exponent $\mu\approx0.5$ 
in zero magnetic field for doped semiconductors without impurity 
compensation has been confirmed.  
On the insulating side of the MIT, 
while the relation $2\nu\approx\zeta$ predicted by scaling 
theories~[15] holds for $0.9<N/N_c<1$, 
the critical exponents for localization length 
and impurity dielectric susceptibility change at $N/N_c\approx 0.99$.  
The small amount of doping compensation that is unavoidably present 
in our samples may be responsible for such a change in the exponents.  
We have also measured the conductivity in magnetic fields up to $B=8$~T 
in order to study the doping-induced MIT (in magnetic fields) 
and the magnetic-field-induced MIT.  
For both of the MIT, the critical exponent of the conductivity is 1.1, 
which is different from the value 0.5 at $B=0$.  
The change of the critical exponent caused by the applied magnetic fields 
supports a picture in which $\mu$ varies depending on the universality class 
to which the system belongs.  
The phase diagram has been determined in magnetic 
fields for the $^{70}$Ge:Ga system.

\HEAD{Acknowledgments}
This work has been performed in collaboration with K. M. Itoh, 
Y. Ootuka, and E. E. Haller.  We are thankful to T. Ohtsuki 
for fruitful discussions, and to S. Katsumoto, B. I. Shklovskii, 
M. P. Sarachik, and J. C. Phillips for valuable comments.  
The author is supported by Research Fellowship of 
Japan Society for the Promotion of Science for Young Scientists.

\HEAD{References}
\begin{biblio}
\ref{1}{Lee,~P.A. and Ramakrishnan,~T.V. (1985) Disordered 
electronic systems, {\it Rev. Mod. Phys.} {\bf 57}, 287-337.} 
\ref{2}{Belitz,~D. and Kirkpatrick,~T.R. (1994) The Anderson-Mott 
transition, {\it Rev. Mod. Phys.} {\bf 66}, 261-380.} 
\ref{3}{Wegner,~F.J. (1976) Electrons in disordered systems. 
Scaling near the mobility edge, {\it Z. Phys. B} {\bf 25}, 327-337; 
(1979) The mobility edge problem: continuous symmetry and a conjecture, 
{\it ibid.} {\bf 35}, 207-210.}
\ref{4}{Rosenbaum,~T.F. Milligan,~R.F. Paalanen,~M.A. 
Thomas,~G.A. Bhatt,~R.N. and Lin,~W. (1983) Metal-insulator 
transition in a doped semiconductor, {\it Phys. Rev. B} 
{\bf 27}, 7509-7523.}
\ref{5}{Stupp,~H. Hornung,~M. Lakner,~M. Madel,~O. and 
L\"{o}hneysen,~H.v. (1993) Possible solution of the conductivity 
critical exponent puzzle for the metal-insulator transition in 
heavily doped uncompensated semiconductors, {\it Phys. Rev. Lett.} 
{\bf 71}, 2634-2637.} 
\ref{6}{Waffenschmidt,~S. Pfleiderer,~C. and L\"ohneysen,~H.v. 
(1999) Critical behavior of the conductivity of Si:P at the 
metal-insulator transition under uniaxial stress, {\it Phys. Rev. 
Lett.} {\bf 83}, 3005-3008.}  
\ref{7}{Rosenbaum,~T.F. Thomas,~G.A. and Paalanen,~M.A. (1994) 
Critical behavior of Si:P at the metal-insulator transition, {\it 
Phys. Rev. Lett.} {\bf 72}, 2121.} 
\ref{8}{Zulehner,~W. (1989) Czochralski growth of silicon, in 
Harbeke,~G. and Schulz,~M.J. (eds.) {\it Semiconductor Silicon: 
Material Science and Technology}, Springer-Verlag, Berlin, 
pp.~2-23.}
\ref{9}{Haller,~E.E. Palaio,~N.P. Rodder,~M. Hansen,~W.L. and 
Kreysa,~E. (1984) NTD germanium: a novel material for low temperature 
bolometers, in Larrabee,~R.D. (ed.) {\it Neutron Transmutation Doping of 
Semiconductor Materials}, Plenum, New York, pp.~21--36.}  
\ref{10}{Itoh,~K.M. Haller,~E.E. Beeman,~J.W. Hansen,~W.L. Emes,~J. 
Reichertz,~L.A. Kreysa,~E. Shutt,~T. Cummings,~A. Stockwell,~W. 
Sadoulet,~B. Muto,~J. Farmer,~J.W. and Ozhogin,~V.I. (1996) 
Hopping conduction and metal-insulator transition in isotopically 
enriched neutron-transmutation-doped $^{70}$Ge:Ga, {\it Phys. Rev. 
Lett.} {\bf 77}, 4058-4061.}
\ref{11}{Watanabe,~M. Ootuka,~Y. Itoh,~K.M. and Haller,~E.E. (1998)
Electrical properties of isotopically enriched neutron-transmutation-doped 
$^{70}$Ge:Ga near the metal-insulator transition, {\it Phys. Rev. B} 
{\bf 58}, 9851-9857.}
\ref{12}{Watanabe,~M. Itoh,~K.M. Ootuka,~Y. and Haller,~E.E. (2000) 
Localization length and impurity dielectric susceptibility in the 
critical regime of the metal-insulator transition in homogeneously 
doped {\it p}-type Ge, {\it Phys. Rev. B} {\bf 62}, R2255-R2258.}
\ref{13}{Shklovskii,~B.I. and Efros,~A.L. (1984) 
{\it Electronic Properties of Doped Semiconductors},  
Springer-Verlag, Berlin.}  
\ref{14}{Ionov,~A.N. Shlimak~I.S. and Matveev,~M.N. (1983) An 
experimental determination of the critical exponents at the 
metal-insulator transition, {\it Solid State Commun.} {\bf 47}, 
763-766.}  
\ref{15}{Kawabata,~A. (1984) Renormalization group theory 
of metal-insulator transition in doped silicon, {\it J. Phys. 
Soc. Jpn.} {\bf 53}, 318-323.}  
\ref{16}{Hess,~H.F. DeConde,~K. Rosenbaum,~T.F. and Thomas,~G.A. 
(1982) Giant dielectric constants at the approach to the 
insulator-metal transition, {\it Phys. Rev. B} {\bf 25}, 5578-5580.}
\ref{17}{Katsumoto,~S. (1990) Photo-induced metal-insulator 
transition in a semiconductor, in Kuchar,~F. Heinrich,~H. and 
Bauer,~G. (eds.) {\it Localization and Confinement of Electrons 
in Semiconductors}, Springer-Verlag, Berlin, pp.~117-126.}
\ref{18}{Chayes,~J.T. Chayes,~L. Fisher,~D.S. and Spencer,~T. 
(1986) Finite-size scaling and correlation lengths for disordered 
systems, {\it Phys. Rev. Lett.} {\bf 57}, 2999-3002.}
\ref{19}{Kirkpatrick,~T.R. and Belitz,~D. (1993) Logarithmic 
corrections to scaling near the metal-insulator transition, {\it 
Phys. Rev. Lett.} {\bf 70}, 974-977.}
\ref{20}{Watanabe,~M. Itoh,~K.M. Ootuka,~Y. and Haller,~E.E. (1999) 
Metal-insulator transition of isotopically enriched 
neutron-transmutation-doped $^{70}$Ge:Ga in magnetic fields, 
{\it Phys. Rev. B} {\bf 62}, 15817-15823.}
\ref{21}{Al'tshuler,~B.L. and Aronov,~A.G. (1983) 
Scaling theory of Anderson's transition for interacting 
electrons, {\it JETP Lett.} {\bf 37}, 410-413.}
\ref{22}{Ohtsuki,~T. and Kawarabayashi,~T. (1997) Anomalous 
diffusion at the Anderson transitions, {\it J. Phys. Soc. 
Jpn.} {\bf 66}, 314-317.}
\ref{23}{Abrahams,~E. Anderson,~P.W. Licciardello,~D.C. and
Ramakrishnan,~T.V. (1979) Scaling theory of localization: 
absence of quantum diffusion in two dimensions, {\it 
Phys. Rev. Lett.} {\bf 42}, 673-676.}
\ref{24}{Bernreuther,~W. and Wegner,~F.J. (1986) Four-loop-order 
$\beta$~function for two-dimensional nonlinear sigma models, 
{\it Phys. Rev. Lett.} {\bf 57}, 1383-1385.}  
\ref{25}{Castner,~T.G. Lee,~N.K. Tan,~H.S. Moberly,~L. and Symko~O. 
(1980) The low-frequency, low-temperature dielectric behavior of 
{\it n}-type germanium below the insulator-metal transition, 
{\it J. Low Temp. Phys.} {\bf 38}, 447-473.}
\ref{26}{Bhatt,~R.N. and Rice,~T.M. (1980) Clustering in the approach 
to the metal-insulator transition, {\it Philos. Mag. B} {\bf 42}, 
859-872.}
\ref{27}{Rentzsch,~R. Ionov,~A.N. Reich,~Ch. M\"uller,~M. Sandow,~B. 
Fozooni,~P. Lea,~M.J. Ginodman,~V. and Shlimak,~I. (1998) 
The scaling behaviour of the metal-insulator transition 
of isotopically engineered neutron-transmutation doped germanium, 
{\it Phys. Status Solidi B} {\bf 205}, 269-273.}  
\ref{28}{Itoh,~K.M. Watanabe,~M. Ootuka,~Y. and Haller,~E.E. (1999) 
Scaling analysis of the low temperature conductivity in 
neutron-transmutation-doped $^{70}$Ge:Ga, {\it Ann. Phys. (Leipzig)} 
{\bf 8}, 631-637.}  
\ref{29}{Khmel'nitskii,~D.E. and Larkin,~A.I. (1981) Mobility edge shift 
in external magnetic field, {\it Solid State Commun.} {\bf 39}, 1069-1070.}
\end{biblio}

\end{document}